\documentclass{article}

\usepackage{microtype}
\usepackage{graphicx}
\usepackage{subfigure}
\usepackage{booktabs} 

\usepackage{hyperref}
\usepackage{subcaption}
\usepackage{caption}

\usepackage[accepted]{icml2024}

\usepackage{amsmath}
\usepackage{amssymb}
\usepackage{mathtools}
\usepackage{amsthm}
\usepackage{wrapfig}
\usepackage[capitalize,noabbrev]{cleveref}

\theoremstyle{plain}

\theoremstyle{definition}

\theoremstyle{remark}

\usepackage[textsize=tiny]{todonotes}

\icmltitlerunning{Teaching dark matter simulations to speak the halo language}

\begin{document}

\twocolumn[
\icmltitle{Teaching dark matter simulations to speak the halo language}




\begin{icmlauthorlist}
\icmlauthor{Shivam Pandey}{yyy}
\icmlauthor{Francois Lanusse}{xxx}
\icmlauthor{Chirag Modi}{xxx}
\icmlauthor{Benjamin D. Wandelt}{xxx,zzz}
\end{icmlauthorlist}

\icmlaffiliation{yyy}{Columbia Astrophysics Laboratory, Columbia University, 550 West 120th Street, New York, NY 10027, USA}
\icmlaffiliation{xxx}{The Flatiron Institute, 162 5th Ave, New York, NY, 10010, USA}
\icmlaffiliation{zzz}{CNRS \& Sorbonne Université, Institut d’Astrophysique de Paris (IAP), Paris, France}

\icmlcorrespondingauthor{Shivam Pandey}{sp4204@columbia.edu}

\icmlkeywords{Machine Learning, ICML}

\vskip 0.3in
]

\printAffiliationsAndNotice{} 

\begin{abstract}
We develop a transformer-based conditional generative model for discrete point objects and their properties. We use it to build a model for populating cosmological simulations with gravitationally collapsed structures called dark matter halos.
Specifically, we condition our model with dark matter distribution obtained from fast, approximate simulations to recover the correct three-dimensional positions and masses of individual halos. 
This leads to a first model that can recover the statistical properties of the halos at small scales to better than 3\% level using an accelerated dark matter simulation. This trained model can then be applied to simulations with significantly larger volumes which would otherwise be computationally prohibitive with traditional simulations, and also provides a crucial missing link in making end-to-end differentiable cosmological simulations. The code, named GOTHAM (Generative cOnditional Transformer for Halo's Auto-regressive Modeling) is publicly available at \url{https://github.com/shivampcosmo/GOTHAM}.
\end{abstract}

\section{Introduction}\label{sec:intro}
Transformer-based architecture, that is now a staple in natural language processing (NLP) applications, excels at learning the conditional distribution of data. It scales well, is highly flexible, and has a native auto-regressive structure. Through the attention mechanism, it can learn the syntactical meaning of a token in relation to other surrounding tokens. This property has remarkable applications beyond NLP. Here we apply the transformer-based architecture to solve one of the long-standing cosmological problems: creating a differentiable end-to-end simulations that can be used as a Bayesian forward model to exhaust the information content in the observations. However, the framework and code developed here can in general be applied to any point cloud generation problem that is conditioned on an external continuous field.

Cosmological N-body simulations evolve a system of more than a billion dark matter particles under gravity from initial Gaussian random fluctuations to present state non-Gaussian large-scale structure. After billions of years of evolution, most of the mass in the simulations ends up in the collapsed dark matter structure called halos. These form the hosts of the galaxies we observe and are, to first-order, described by their masses. However, solving particle-particle interactions of billions of particles for many time steps makes these simulations computationally intensive. Moreover, finding and characterizing the collapsed dark matter structures further exacerbates the computational requirements. Finally, both the traditional N-body simulations and halo finders are non-differentiable processes, making it challenging to use machine learning-based methods for their analysis.

Particle mesh (PM) simulations instead put all the particles on a regular grid and solve their equations of motion using fast Fourier transform techniques \citep{Feng:2016:MNRAS:}. These simulations capture the large-scale dark matter distribution accurately, are significantly faster compared to N-body, and can also be written in a differentiable form \citep{Modi:2021:A&C:, Li2022}. However, as the particles are placed on a grid for PM simulation, it lacks the resolution capabilities of an N-body simulation and underestimates the small-scale structures. As the dark matter halos form and get their properties from small-scale interaction of the particles, this means that the PM simulations severely underestimate the number and properties of dark matter halos. In this study, we learn the mapping between large-scale dark matter distribution as obtained from PM simulations and N-body-like dark matter halos. This architecture can also be made differentiable \citep{Horowitz:2024:MNRAS:}, which when combined with PM simulations, leads to an end-to-end differentiable cosmological simulator.

Having an accurate differentiable forward model is required if we hope to extract most of the information in the observed galaxy data. The volume of the galaxy survey \citep{Alam_2015} that ended a decade ago is at least a factor of 27 larger than available from the current best N-body simulation suites \citep{Villaescusa_Navarro_2020} for forward modeling the observations. The model developed here describes a method to use fast and approximate dark matter simulations that are easy to scale to obtain N-body-like halo catalogs.

\section{Related Works}\label{sec:related_works}
As halos are a set of discrete objects in 3D space, techniques based on point cloud and their transformations are relevant. Several studies have implemented various ways of completing a point cloud by leveraging the features learned from the given partial set of points \citep{yan2022shapeformer, yu2022pointbert}. 
However, here we are interested in generating a full set of point clouds conditioned only on the surrounding dark matter distribution which requires a different architecture. Moreover, usually, the point cloud transformers treat each point as unweighted. However, for halos, their masses are an important property to capture accurately as they can significantly impact the properties of galaxies it hosts which is what we measure in the observations. Therefore, here we develop an architecture that can jointly infer the position and masses of the halos. 

Several past studies have also aimed at learning a mapping from simulated dark matter distributions to dark matter halos. For example, \citet{Ramanah:2019PhRvD.100d3515K} trained a Wasserstein GAN to predict counts of halos within four mass bins based on gridded dark matter density from a full N-body simulation. \citet{jamieson2022} learn the correction to velocities and positions of the particles in PM simulations to replicate N-body-like simulations. However, even after the correction, the statistical properties of halos are only reproduced at 20-30\% level. \citet{charm_mnras} learn the mapping between PM simulation and halos but are limited to large scales ($k < 0.3 h/$Mpc) in reproducing the statistical properties of halos. In this work, we treat the conditional halo catalog generation as a language translation problem and use a large language model-like architecture to solve it. The flexibility of such a model allows us to model the halo distribution and their masses at the 2-3\% level to significantly smaller scales ($k \sim 1 h$/Mpc).

\begin{figure*}[ht]
\centering
\includegraphics[width=1.\textwidth, height=0.55\textwidth]{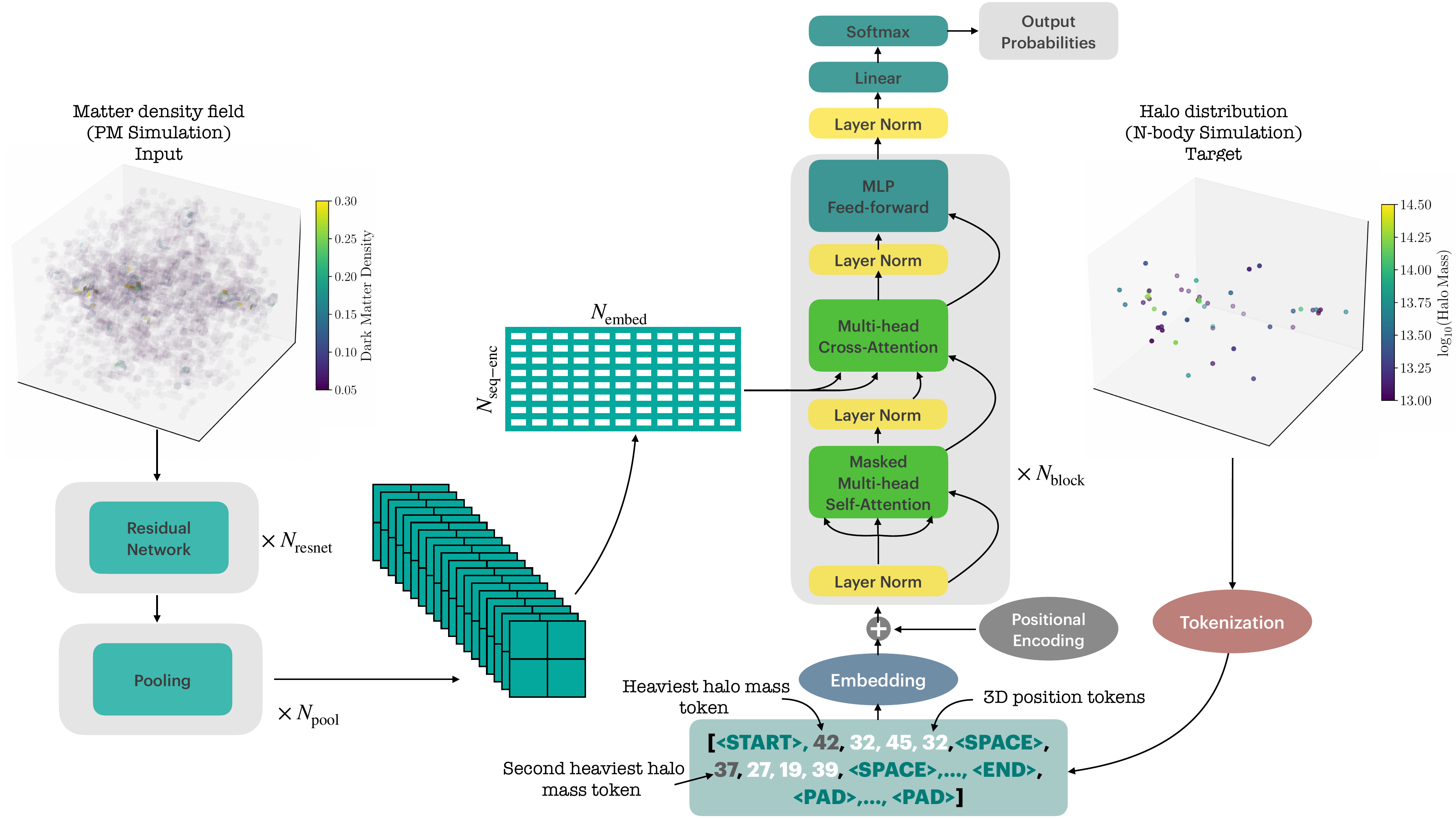} 
\caption{Model architecture: On the left we show the input dark matter density distribution for one training sub-box and on the right we show the target 3D distribution of halos colored by their masses. These four properties are tokenized and concatenated for all the halos in the sub-box in a way that it forms a `sentence'. On the encoder side, we use stacks of 3D residual networks to extract the features from the density field and input them to the cross-attention module of the decoder model to learn the conditional log probability of the tokens. See Sec.~\ref{sec:dataset} and Sec.~\ref{sec:methods} for more details.}
\label{fig:architecture}
\end{figure*}

\section{Dataset}\label{sec:dataset}
 The halo catalogs for training are obtained from the public Quijote N-body `high-resolution' simulation suite at a fiducial cosmology \citep{Villaescusa_Navarro_2020} which evolve $1024^3$ particles inside a box with a side length of 1000 Mpc/$h$. We learn the distribution of these catalogs when conditioned on 3D dark matter densities derived from the low-resolution PM simulations. 

\subsection{Input : Continuous 3D conditional field}
We run PM simulations with same initial conditions and volume as the Quijote simulations. However, to significantly reduce the run time, we run them with only $384^3$ particles and the forces between particles are calculated in a $768^3$ grid. When run on CPUs, each simulation has a runtime of 5 CPU hours (c.f. 5000 CPU hours for N-body simulation), which can be further reduced by using its GPU implementation \citep{Li2022}. We run 11 different PM simulations for different set of initial conditions to capture the stochastic contribution. 
We sub-divide each parent 1000 Mpc/$h$ box into $32^3$ sub-boxes (giving $L_{\rm sub-box}=31.25 \, {\rm Mpc}/h$) and treat each of these sub-boxes as independent. We use the sub-boxes from first three simulations for training (80\%) and validation (20\%) and use remaining eight simulations as test set.

\subsection{Target: Weighted discrete 3D point cloud}
We use an accurate definition of halos which uses phase space distribution of dark matter particles to identify collapsed structures and assign them masses \citep{Behroozi_2012}. In this study, we only focus on halos with masses above $M_{\rm halo} > 10^{13.5} M_{\odot}/h$. We aim to learn the three spatial coordinates and mass of each halo, when conditioned on the input density field.

\subsection{Tokenization}\label{sec:token}
As the position and masses of the halos are continuous, we tokenize them. First we scale each of 3D coordinate and logarithm of masses in the range $(0, 1)$. Then we divide them into 64 bins and assign each halo four integers corresponding to its 3D coordinate and mass. Therefore, each halo effectively becomes a ``word" with four characters. We concatenate the tokens of all the halos in the sub-box by spacing them with a $\texttt{<SPACE>}$ token. Finally we pre-pend a $\texttt{<START>}$ token and append an $\texttt{<END>}$ token to create a ``sentence" of halos for each sub-box (giving $n_{\rm vocab} = 67$). As we physically expect the halo with heaviest mass to dominate the structure formation in a sub-box, we concatenate the halos with a descending order of their masses. This is then padded with a $\texttt{<PAD>}$  token to reach the maximum sequence length of $N_{\rm seq-dec}=101$ (corresponding to having a maximum of 20 halos in a sub-box). 
We then create a right-shifted prediction vector to predict these tokens conditioned on the input field. Therefore the task effectively becomes understanding the ``grammatical" structure of these halo ``sentences".

\begin{figure*}[htb!]

    \begin{subfigure}
        \centering
        \includegraphics[width=1.\textwidth, height=0.375\textwidth]{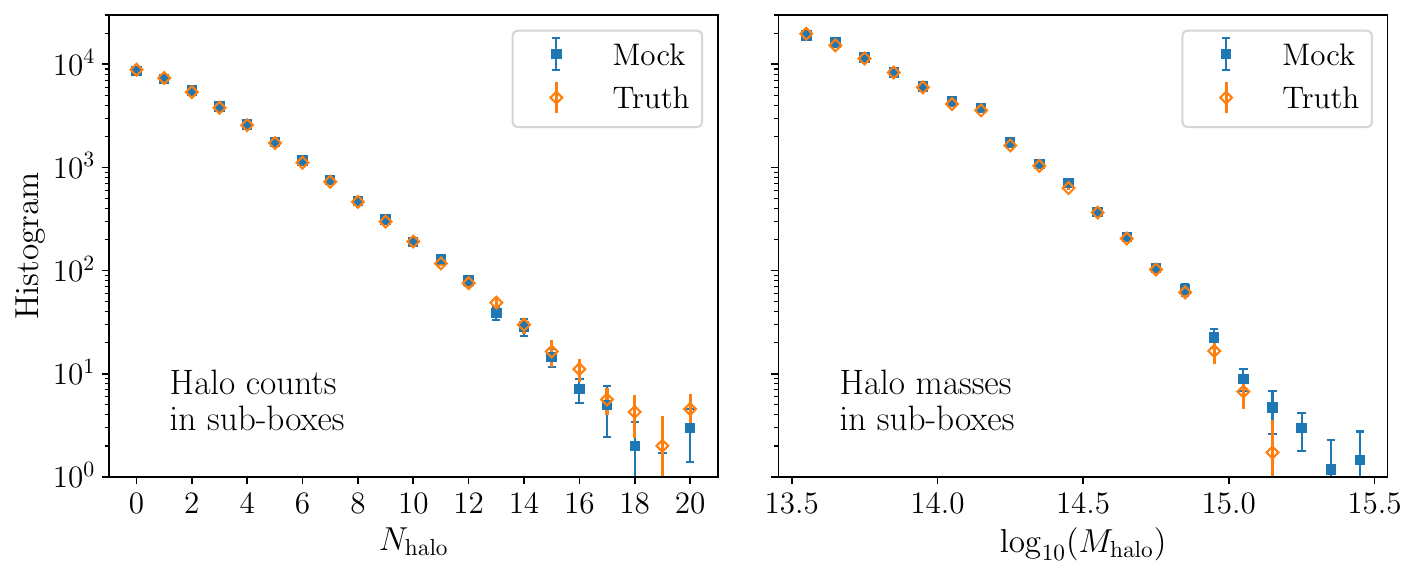}
        \label{fig:halo_counts_masses}
    \end{subfigure}

    \vskip -0.2in

    \begin{subfigure}
        \centering
        \includegraphics[width=1.\textwidth, height=0.375\textwidth]{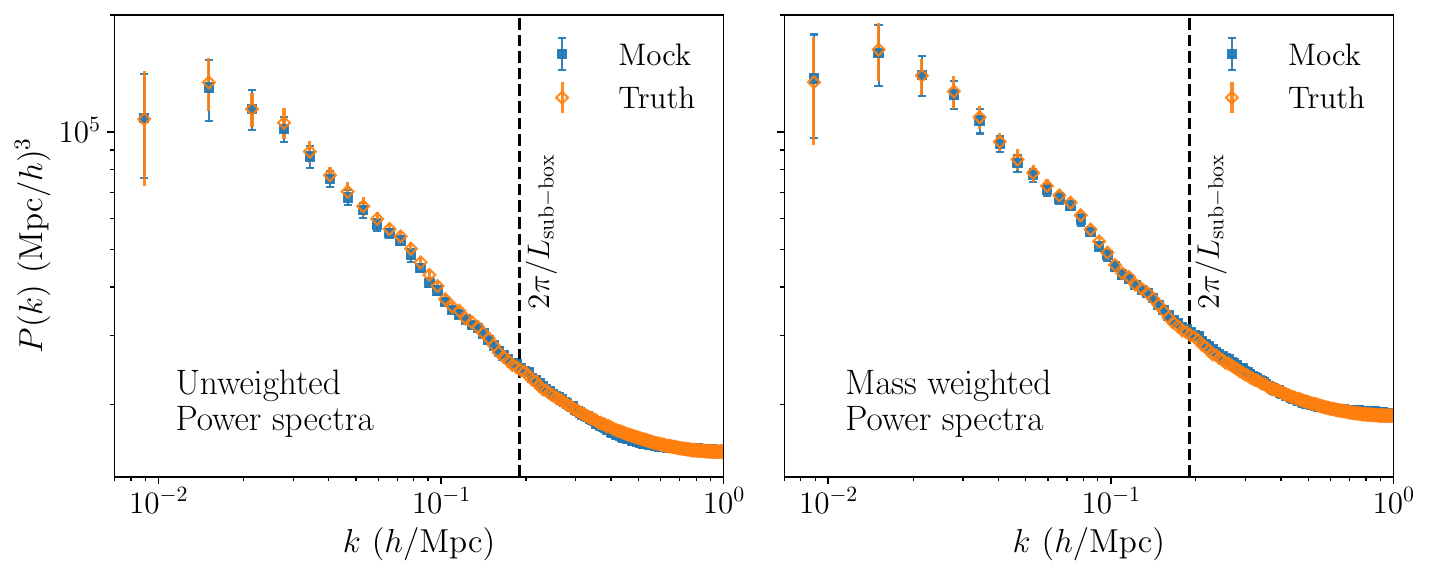}
        \label{fig:halo_power_spectra}
    \end{subfigure}
    \vskip -0.35in
    \caption{
    \textit{{Top row, local performance}}: Histogram of the predicted and true halo number counts (left) and masses (right) in 8 test simulations. We also show the standard deviation in true and mock halo catalogs, finding that the trained network can accurately capture the mean and their uncertainties.
    \textit{Bottom row, global performance}:  Comparison of the power spectrum of the mock and true halo catalogs, unweighted (left) or weighted by mass (right). 
    We also show the Nyquist frequency of the sub-box ($L_{\rm sub-box} = 31.25 \, {\rm Mpc}/h$). 
    }
    \label{fig:combined_performance}
\end{figure*}

\section{Methodology}\label{sec:methods}
We use the transformer architecture along with residual network in this task. Similar to NLP models \citep{vaswani2023attention}, we can divide our architecture into two parts, encoder and decoder. 
We show the details of the architecture in Fig.~\ref{fig:architecture}.

\subsection{Encoder}
To extract the feature vectors that can inform the halo selection, we run stack of $N_{\rm res-net}=4$ residual networks. We input three PM density fields to the network, each with a resolution of $32 \times 32 \times 32$, but obtained from a physical volume of $32$, $48$ and $96$ Mpc$/h$ respectively, centered on each sub-box. This captures the information from the surrounding environment that is crucial to capture halo formation physics. After passing through the residual network with a filter of size $n_f = 3$, we obtain an output with shape $16 \times 16 \times 16 \times N_{\rm embed}$, where $N_{\rm embed}=64$ are the features we extract. We then spatially downsample the output with $N_{\rm pool} = 4$ layers to get the final output with shape of $4 \times 4 \times 4  \times N_{\rm embed}$. This output is spatially flattened to finally obtain a matrix of shape $N_{\rm seq-enc} \times N_{\rm embed}$, where $N_{\rm seq-enc} = 64$ and this is used as input to the cross-attention in the decoder module.

\subsection{Decoder}
The decoder part mostly follows the architecture introduced in \citet{vaswani2023attention}. The tokenized halo ``sentence" (Sec.~\ref{sec:token}) is first embedded to $N_{\rm embed} = 64$ dimensions.  We then add linear positional embedding to this input and pass it to a stack of $N_{\rm block} = 4$ multi-head attention modules, each with $N_{\rm head}=4$ heads. Note that the cross-attention module gets its key and value from the encoder described above whereas the query is generated from the halo tokens. We use standard multi-class cross-entropy loss over $n_{\rm vocab}=67$ classes to predict the next token number, conditioned on previous tokens and the features from the PM density fields. 

When predicting the mock halo catalog from test simulations, we provide the encoder the dark matter density from the PM simulations and a $\texttt{<START>}$ token to the decoder part. We end the prediction once the $\texttt{<END>}$ token is predicted. The predicted token numbers are then used to convert back to the 3D positions within the sub-box and the masses of the halos.

\section{Results}\label{sec:results}
\subsection{Local performance}
We compare the histogram of the predicted mock halo catalogs to the true N-body catalogs on 8 test simulations in the top panel of Fig.~\ref{fig:combined_performance}. We find that the architecture can correctly predict the number of the halos in each sub-box as well as their masses. We show the mean of the histogram obtained from 8 simulations as well as their standard deviation, finding that the stochastic uncertainties due to varying initial conditions are also correctly captured in the model. This tests the local performance of the model.

\subsection{Global performance}
To test the global performance of the model, we take the predicted mock catalogs in each sub-box (of box-length, $L_{\rm sub-box} = 31.25 \, {\rm Mpc}/h$) corresponding to one realization and re-create the distribution in the full box of length $L_{\rm full-box} = 1000 \, {\rm Mpc}/h$ by stacking the sub-volumes. We then measure the power spectrum of the halos in this full box on a wide range of scales as shown in the bottom left panel of Fig.~\ref{fig:combined_performance}. We find that the resulting mock power spectrum matches to the true power spectrum at 3\% level. To test that the model has correctly captured the correlation between halo masses and their spatial distribution, we additionally weight each halo with their predicted mass value using a power-law weighting ($w_i = (M/M_1)^{\alpha}$, where $M_1 = 10^{14}M_{\odot}/h$ and $\alpha=0.66$) and calculate the power spectrum. The comparison is shown in the bottom right panel of Fig.~\ref{fig:combined_performance} and we again find a 3\% agreement between mock and truth catalogs. We show the mean and variance of the prediction and true power spectra calculated from 8 test simulations and we find that the mock catalogs also correctly capture the variance in the power. Note that as the power matches well between true and mock catalogs on both the scales much smaller and larger compared to the size of sub-boxes ($k \sim 2\pi/L_{\rm sub-box} = 0.2 \, h/{\rm Mpc}$), the model is capturing the correlation between spatial position and masses correctly.

\section{Conclusion}\label{sec:conclusion}
In this study we developed a novel model to generate conditional distribution of weighted discrete point cloud (dark matter halos) when conditioned on a correlated 3D field (dark matter density field). We showed that such a model can accurately reproduce the catalog such that it has correct distribution and statistics. We plan to enhance the network to correctly predict even lower mass halos and include more properties in the inference such as velocity and concentration of the dark matter halo by increasing the size of each ``word" in the halo ``sentence" (see Sec.~\ref{sec:token}). We also plan to generalize the model for different cosmologies. The code is publicly available at  \url{https://github.com/shivampcosmo/GOTHAM}

\section*{Acknowledgments}
SP thanks the CCA and CCM at the Flatiron Institute for hospitality while (a portion of) this research was carried out. The computations reported in this paper were performed using resources made available by the Flatiron Institute. The Flatiron Institute is supported by the Simons Foundation. 
This work is supported by the Simons Collaboration on ``Learning the Universe''. 

\bibliography{paper}
\bibliographystyle{icml2024}

\end{document}